\documentstyle[11pt,aaspp4]{article}

\lefthead{Taniguchi et al.}
\righthead{The most distant [OIII] emitting quasar PKS 1937$-$101}

\begin{document}

\title{The most Distant [OIII]-emitting Quasar PKS 1937$-$101 at redshift 3.8}

\author{Yoshiaki Taniguchi\altaffilmark{1}}
\affil{Astronomical Institute, Tohoku University, Aoba, Sendai 980-77, Japan 
and Royal Greenwich Observatory, Madingley Road, Cambridge CB3 0EZ, UK;
tani@astroa.astr.tohoku.ac.jp}
\author{Takashi Murayama\altaffilmark{1}}
\affil{Astronomical Institute, Tohoku University, Aoba, Sendai 980-77, Japan;
murayama@astroa.astr.tohoku.ac.jp}
\author{Kimiaki Kawara\altaffilmark{1}}
\affil{Institute of Astronomy, The University of Tokyo, 2-21-1 Osawa,
Mitaka, Tokyo 181, Japan; kkawara@mtk.ioa.s.u-tokyo.ac.jp}
\author{Nobuo Arimoto}
\affil{Institute of Astronomy, The University of Tokyo, 2-21-1 Osawa,
Mitaka, Tokyo, 181 Japan
and Institute of Astronomy, University of Cambridge, Madingley Road, Cambridge
CB3 0HA, UK;  Physics Department, University of Durham,
South Road, Durham, DH1 3LE, UK; arimoto@mtk.ioa.s.u-tokyo.ac.jp}

\author{(Received 10 April 1997;\hspace*{1em}accepted 30 April 1997)}
\author{To appear in Publications of the Astronomical Society of Japan}

\altaffiltext{1}{Visiting astronomer, Kitt Peak National Observatory, National 
Optical Astronomy Observatories, which are operated by the Association of 
Universities for Research in Astronomy, Inc., under contract with the National
Science Foundation.}


\begin{abstract}

We report the discovery of a  high-$z$ quasar
with unambiguous [OIII]$\lambda$5007 emission; PKS 1937$-$101 at
redshift 3.8. This quasar, however, shows  
little evidence for rest-frame ultraviolet and optical FeII emission.
It is thus shown that PKS 1937$-$101 does not belong to a class
of super iron-rich high-$z$ quasars reported by Elston, Thompson, \& Hill (1994).
The epoch of major star formation in the host galaxy is discussed briefly.

\end{abstract}


\section{Introduction}

The major epoch of star formation in galaxies is one of the
most important topics in modern astrophysics,
because it is significantly related to the formation of galaxies
and quasars as well as to cosmology.
Massive stars formed in the first episode of
star formation have a lifetime of
$10^6$ to $10^7$ years
and then  release Type II supernova (SNII) products
(primarily the $\alpha$-elements such as O, Ne, Mg, Si, etc., but
comparatively little iron).
It takes a much longer time for
Type Ia supernovae (SNIa)
to release  iron.
The different nucleosynthesis yields and timescales of SNIa's
and SNII's thus make the abundance ratio [$\alpha$/Fe]
a potentially useful cosmological clock with which one
can identify the epoch of first star formation in galaxies.
It is therefore important to
study chemical properties of high-redshift ($z$) objects.
The best objects are, however, not galaxies but quasars.
Quasars possess broad emission-line gas in the nuclear region,
which is photoionized by the central black hole engine.
It is generally considered that the heavy elements in the broad line regions (BLRs)
come from stars in a host galaxy.
Therefore, systematic study of chemical properties of BLRs
of quasars at high redshift is of particular interest
(Hamann \& Ferland 1992, 1993).
Rest-frame optical emission lines, which are usually used to study chemical
properties of nearby objects,  are redshifted to the near-infrared (NIR)
in these quasars.
Recent NIR spectroscopy of high-$z$ quasars has shown that
the rest-frame optical spectra are dominated by singly ionized iron (FeII) emission
as well as
hydrogen recombination lines (Hill, Thompson, \& Elston 1993;
Elston, Thompson, \& Hill 1994; Kawara et al. 1996),
suggesting long-lasting star formation 
in the nuclear regions of the quasar hosts ($\sim 1$ Gyr).

Recently Kawara et al. (1996) detected [OIII]$\lambda$5007 emission 
from the high-$z$ quasar B1422+231 at redshift 3.62,
and showed that the rest-frame UV and optical spectrum is quite similar to 
the average spectrum of LBQS (Large Bright Quasar Survey) quasars which are mostly
located at $z \sim$ 1 - 2 (Francis et al. 1991).
However, most high-$z$ quasars studied by Hill et al. (1993) and Elston et al. 
(1994) turn out to be super iron-rich quasars which are not so frequently
observed in nearby quasars (cf. L\'ipari, Terlevich, \& Macchetto 1993).
In order to investigate star formation history of high-$z$ quasar hosts,
we need more NIR spectroscopic observations of high-$z$ quasars
(cf. Taniguchi et al. 1996).
Here, we report our new NIR spectroscopy of  a  high-$z$ quasar
PKS 1937$-$101,
which  is a radio-loud quasar with a flat spectrum at a redshift of 3.787
(Bolton, Savage, \& Wright 1979; Lanzetta et al. 1991).
Despite of its high redshift,
its optical apparent magnitude is as bright as $V$ = 17.0 mag 
(V\'eron-Cetty \& V\'eron 1996)
and so the absolute magnitude is $M_V = -28.8$ mag
(a Hubble constant $H_0$ = 100 km s$^{-1}$ Mpc$^{-1}$, and a deceleration
parameter $q_0 = 0$ are assumed). The soft X-ray luminosity
is also huge, $L_{\rm X} \sim 2\times 10^{47}$ erg s$^{-1}$ 
(Brinkmann \& Siebert 1995).
Hence, PKS 1937$-$101 is one of the most intrinsically luminous quasars  
ever known.

\section{Observations and Data Reductions}

PKS 1937$-$101 was observed by using the long-slit
Cryogenic Spectrometer (CRSP: Joyce 1995)
 with a $256 \times 256$ InSb detector array at the f/15 focus of
the Kitt Peak National Observatory
 (KPNO) 4 meter telescope on 17 May 1995.
A $2.3'' \times 49''$ slit with a scale of 0.36$''$ pixel$^{-1}$ was placed
on the intensity peak of the object in the EW direction.
The seeing size was 2.7$''$.
A 200 lines mm$^{-1}$
grating with a blaze angle of $17.5^\circ$ was used
in second order at
$J$ ($1.095 {\rm -} 1.349\ \mu$m),
and in first order at $K$ ($2.055 {\rm -} 2.430\ \mu$m). OH
airglow lines and HeNeAr lamp spectra were used to calibrate the wavelength
scale and to measure the spectral resolution.  The accuracy of the wavelength
scale calibration is 130 km s$^{-1}$ at $J$,
and 110 km s$^{-1}$ at $K$.  The typical FWHMs of the spectral resolution are
2000 km s$^{-1}$ at $J$ and 2300 km s$^{-1}$ at $K$.
The object was  shifted along the slit by 10 arcsec between
exposures.
The redshift of PKS 1937$-$101 is so high ($z = 3.8$) that the conspicuous
rest-frame optical emission lines ([OIII]$\lambda5007$, H$\beta$, and some FeII)
are redshifted
beyond 2.3 $\mu$m. Although the $K$-band spectral coverage extends to 2.43 $\mu$m,
 the background
thermal emission makes it difficult to observe at $\lambda >$ 2.3 $\mu$m.
We therefore set the unit integration time of $K$-band observations as 20 s to
avoid the background saturation,
and took 120 scans to achieve the total integration time = 2400 s.
Another interesting emission line MgII$\lambda$2798 is slightly 
out of the observable range of $J$ band
but we made $J$-band spectroscopy of PKS 1937$-$101 to see
the ultraviolet FeII features as well as the continuum emission.
Because the observing time was limited, we took only two 300 s data of
$J$ band (the total integration time = 600 s).
The data reduction was done using standard techniques (Joyce 1995).
The residual sky
emission was removed by fitting the sky emission on  adjacent sky pixels.
A faint standard star HD 162208 (A0; $J$=7.215 and $K$= 7.110) was
used to calibrate the 
flux scale and to correct for  telluric extinction.  A 10000 K black-body
spectrum, which fits to the $JHKL$ magnitudes
of the standard star (Elias et al. 1982) within
5\% deviation, was used for flux scale calibration.
The photometric accuracy is estimated to be 7 \% in $J$ and 13 \% in $K$.

\section{Results and Discussion}

Figure 1 shows the observed $J$- and $K$- band spectra of PKS 1937$-$101
together with the LBQS
composite spectrum of quasars at $z = 1 {\rm -} 2$
(Francis et al. 1991).
The expected FeII emission features are also shown in the lower panel
for reference.
The $K$-band spectrum  shows [OIII]$\lambda$5007, H$\beta$, 
and H$\gamma$ ([OIII]$\lambda$4363
possibly overlaps with this line).
The emission-line properties are summarized in Table 1.
The H$\beta$ emission seems narrower than those of typical quasars
(see Table 1 and Figure 2).
The [OIII]/H$\beta$ ratio is similar to that of LBQS composite quasar spectrum.
In Figure 2, we show that the observed $K$-band spectrum can be 
fitted well solely by
the emission lines of [OIII]$\lambda$4959,5007, H$\beta$, H$\gamma$,
and the linear continuum.
This fitting gives FeII(4434-4686)/H$\beta \simeq 0.1$ nominally, being
smaller than that of B1422+231 (Kawara et al. 1996).
We also show the comparison of the rest-frame optical spectra
between PKS 1937$-$101 and B1422+231 (Kawara et al. 1996) in Fig. 3.
There cannot  be seen strong 
optical FeII emission in the PKS 1937$-$101 spectrum.
However, taking account of the poorer S/N of the PKS 1937$-$101 spectrum,
we give a 3$\sigma_{\rm rms}$ upper limit in Table 1
where $\sigma_{\rm rms}$ is the root mean square noise of flux in the range
between 4434 \AA~ and 4686 \AA.

The $J$-band spectrum shows 
little evidence for ultraviolet FeII emission feature, either.
In the red edge of the $J$-band spectrum,
a blue part of MgII$\lambda$2798 emission can be
seen  although we cannot estimate anything about this line.
Given the photometric accuracy of our $J$- and $K$-band spectra,
the difference of fluxes in the $J$ band
between ours and the LBQS composite spectrum may well be real.
In fact, we can fit the continuum emission with a power law of
$F_\nu \propto \nu^{-0.50}$ (see Figure 1),
which is almost consistent with the average continuum spectrum of quasars,
where the power-law index ranges from $-0.3$
(Francis et al. 1991) to $-0.7$ (Sargent et al. 1989).
The ultraviolet spectra of most quasars,
regardless of radio loudness (Bergeron \& Kunth 1984),
are dominated  by the FeII features as well as
the power-law continuum emission.
Therefore, the lower flux and the featureless property of the $J$ band spectrum
may provide evidence against the presence of  ultraviolet FeII emission features
in PKS 1937$-$101.
We therefore conclude that 
PKS 1937$-$101 does not belong to a class of super 
iron-rich high-$z$ quasars\footnote{Our new measurement of one of the super 
iron-rich high-$z$ quasars studied by Elston et al. (1994) has shown that S4 0680+68
at redshift 3.2 does not belong to this class (Taniguchi et al. 1996;
Murayama et al. 1997).}
reported by Elston et al. (1994).

Our NIR spectroscopy has shown that PKS 1937$-$101 is 
the most distant quasar with [OIII]
emission but little FeII emission. 
Here we discuss the nature of high-$z$ quasars in terms of their rest-frame
optical spectra. We give a summary of the recent 
NIR spectroscopy of high-$z$ quasars including
PKS 1937$-$101 in Table 2.
There is a tendency that the quasars with $z < 3.5$ show strong FeII emission
(Hill et al. 1993; Elston et al. 1994)
while those with $z > 3.5$ show strong [OIII] emission (Kawara et al. 1996; this paper).
One interesting spectroscopic property known for low-$z$ ($z < 0.5$) quasars is 
the anticorrelation between
the strength of optical FeII and [OIII] emission lines,
although its physical mechanism is not fully understood (Boroson \& Green 1992).
We examine if the high-$z$ quasars follow the same anticorrelation.
In Figure 3, we show the relationship of the equivalent width ratios between
([OIII]$\lambda4959+\lambda5007$)/H$\beta$ and FeII$\lambda$4434-4684/H$\beta$.
The low-$z$ quasars studied by Boroson \& Green (1992) show a loose, but
statistically significant  anticorrelation.
It is also shown that the radio-loud quasars tend to be located in the lower portion of
this diagram (i.e., weak FeII emitters). PKS 1937$-$101, B1422+231 (Kawara et al. 1996),
and the radio-quiet, high-$z$ quasars studied by Hill et al. (1993)  
share the same property 
as those of low-$z$ quasars. On the other hand, the radio-loud quasars studied by Elston
et al. (1994) and Hill et al. (1993) do not follow the same trend as
low-$z$ quasars.
We have shown that PKS 1937$-$101 and B1422+231 are members of the class of objects
which share the same
optical emission-line properties as those of low-$z$ quasars.
Since the total number of high-$z$ quasars discussed here is only eight,
we need more observations to understand the general nature of high-$z$ quasars.

Finally we comment on the epoch of major star formation in the host galaxy
of PKS 1937$-$101.
The $\alpha$ elements, such as O and Mg, come from SNII's of massive star origin
and thus are quickly expelled into the interstellar space after
the major episode of star formation
(within a few $10^6$ to 10$^7$ years).
It is considered that the N enrichment is delayed ($\sim 10^8$ years)
because it is partly a secondary element formed by CNO 
cycle in hydrogen-burning shell of intermediate mass stars
(Hamann \& Ferland 1993).
The rest-frame ultraviolet spectra of PKS 1937$-$101 taken by
Lanzetta et al. (1991) and Fang \& Crotts (1995)
show evidence for NV$\lambda$1240 emission.
Therefore, the nuclear gas has already been polluted with N,
implying that the elapsed time from the major star formation is longer than
$\sim 10^8$ years (Hamann \& Ferland 1993).
However, our observation suggests that the major Fe enrichment  has not yet been
made in PKS 1937$-$101.
If this is the case, our observations provide a constraint on
the epoch of major star formation in the host galaxy.
The bulk of iron come from SNIa's whose
progenitors' lifetime is very likely to cluster around $\sim 1.5$ Gyr
(Yoshii, Tsujimoto, \& Nomoto 1996).
Therefore, the Fe enrichment may start at 1.5 Gyr after
the onset of the first, major  star formation in quasar host galaxies.
These arguments, therefore,  specify the epoch of major star formation in PKS 1937$-$101;
$\sim 10^8$ - 1.5$\times 10^9$ years before redshift 3.787.
Namely, the initial star formation would occur
at $ 3.9 < z  < 6.7$ for
$H_0 = 50$ km s$^{-1}$ Mpc$^{-1}$ and $q_0 = 0$, while at $4.0 < z < 17$ for
$H_0 = 100$ km s$^{-1}$ Mpc$^{-1}$ and $q_0 = 0$.
Recent theoretical prescription on the star formation at high-$z$ universe suggests
that the major epoch of star formation may occur $z < 5$ although subgalactic structures may
exist even at $z > 10$ (Rees 1996; Ostriker \& Gnedin 1996).
Provided that the smaller $H_0$ is more preferable,
the present observation is consistent with this prescription.

\acknowledgments

We are very grateful to the staff of KPNO and particularly Dick Joyce for
technical support and assistance with the observations.
We also thank a nice LTO, Dan Yacom for his skillful operation of
the telescope and
picking us to Tucson Airport just after the observation of PKS 1937$-$101.
We would like to thank Martin Rees for discussion about the galaxy formation
at high-$z$ universe, Fred Hamann for  both discussion about chemical properties of
high-$z$ quasars and encouragement, and
Neil Trentham for critical  reading the manuscript and useful comments.
YT also thanks Keith Tritton, Roberto \& Elena Terlevich, and Isabel Salamanca
at Royal Greenwich Observatory for their warm hospitality.
NA is grateful to PPARC for a financial support for his
stay in IoA, University of Cambridge
and Physics Department, University of Durham, 
and would like to thank Richard Ellis,
Alfonso Aragon-Salamanca, Richard Bower, and Roger Davies
 for fruitful discussions.
This work was financially supported in part by Grant-in-Aids for the Scientific
Research (Nos. 06640349, 07222206, and 0704405) of the Japanese Ministry of
Education, Culture, Sport, and Science and by the Foundation for Promotion
of Astronomy, Japan.
TM was supported by the Grant-in-Aid for JSPS Fellows
by the Ministry of Education, Science, Sports, and Culture.

\newpage


\newpage


\figcaption[tani_pks_fig1.eps]{%
The spectrum of PKS 1937$-$101 at $z = 3.787$ in the observed frame.
The composite spectrum of 700 LBQS quasars at $z \sim$ 1 - 2 is also shown.
The dashed curve shows the power-law ($F_\nu \propto \nu^{-0.5}$) continuum
fit for the emission- and absorption-free continua of the
observed $J$- and $K$-band spectra.
The bars at the both sides of the $J$- and $K$-band spectra show the photometric
accuracy.
It is noted that there are two possible absorption features 
at $\lambda$ = 12575 \AA~ and $\lambda = $ 12890 \AA.
If they are attributed to MgII absorbers,
their redshifts would be 3.493 and 3.604, respectively.
The top panel shows the atmospheric transmission curve at the Kitt Peak$^{13}$.
The lower panel shows the expected FeII emission feature for reference.
\label{fig1}}

\figcaption[tani_pks_fig2.eps]{%
The $K$-band spectrum of PKS 1937$-$101. The profile fit is made for
H$\gamma$, H$\beta$, and [OIII]$\lambda$4959,5007.
The H$\beta$ emission is deconvolved into
the narrow (FWHM $\simeq$ 1970 km s$^{-1}$) and  broad  (FWHM $\simeq$ 7710 km s$^{-1}$)
components (see Table 1).
The continuum is fitted linearly. We do not use the data of both
$\lambda_{\rm rest} < 4200$ \AA~ and
$\lambda_{\rm rest} > 5100$ \AA~ in the fitting
because the spectral quality is not good due to the low
atmospheric transmission.
In the lower panel, the residual of the fit is compared with the expected FeII
emission feature.
\label{fig2}}

\figcaption[tani_pks_fig3.eps]{%
The comparison of the rest-frame optical spectra
between  PKS 1937$-$101 and B1422+231 (Kawara et al. 1996).
\label{fig3}}

\figcaption[tani_pks_fig4.eps]{%
Diagram between ([OIII]$\lambda$5007+$\lambda$4959)/H$\beta$
equivalent width ratio
and FeII$\lambda$4434-4684/H$\beta$ one for low-$z$ (small symbols)
and high-$z$ quasars (large symbols).
Radio-quiet, radio-loud with flat spectrum, and radio-loud with steep spectrum
are shown by open circles, filled circles, and filled squares, respectively.
B2 1225+317 is shown by the filled triangle because its radio spectrum is unknown.
The numbers given for the high-$z$ quasars correspond to those in Table 2.
\label{fig4}}


\clearpage

\begin{table*}
\begin{center}
\begin{tabular}{ccccc}
\hline \hline
Line &  $F/F({\rm H}\beta ({\rm N+B}))^a$ & EW(rest)$^b$ & FWHM$^c$ & FWHM$_{\rm cor}^d$  \\
     &  & (\AA) & (km s$^{-1}$) & (km s$^{-1}$)  \\
\hline
H$\gamma$+[OIII]$\lambda$4363 & $0.21 \pm 0.13$ & $11.7 \pm 7.4$ & 6500 & 5900  \\
H$\beta$(N) & $0.46 \pm 0.13$ & $30.5 \pm 8.5$ & 1970 & $<$1970 \\
H$\beta$(B) & $0.54 \pm 0.19$ & $36.1 \pm 13$ & 7710 & 7350  \\
{}[OIII]$\lambda$5007 & $0.27 \pm 0.05$ & $18.8 \pm 4.5$ & 2450 & 960  \\
FeII($\lambda$4434-4684)$^e$ & $<0.49^f$ & $<29^f$ & -- & --  \\
\hline \hline
\end{tabular}
\end{center}


\tablenotetext{a}{
N = narrow component, and B = broad component.
$F({\rm H}\beta ({\rm N+B})) = 3.9 \times 10^{-14}$ erg s$^{-1}$ cm$^{-2}$.}
\tablenotetext{b}{
The rest frame equivalent width.}
\tablenotetext{c}{
Full width at half maximum.}
\tablenotetext{d}{
Full width at half maximum corrected for instrumental broadening.}
\tablenotetext{e}{
To compare with those of low-$z$ quasars studied by Boroson \& Green (1992).}
\tablenotetext{f}{
Upper limit (3 $\sigma$).}
 
\tablenum{1}
\caption{Emission line properties of PKS 1937$-$101 \label{tbl-1}}

\end{table*}


\clearpage

\begin{table*}
\begin{center}
\begin{tabular}{cccccccc}
\hline \hline
No. & Name & Redshift & Type$^a$ & $M_V^b$ & [OIII]/H$\beta^c$ & FeII/H$\beta^d$ & References
$^e$ \\
\hline
1 & B2 1225+317 & 2.219  & Loud  & $-28.5$  & $<$ 0.05 & 1.95 & 1 \\
2 &  Q1246$-$057 & 2.244  & Quiet (BAL) & $-27.6$  & $<$ 0.10 & 1.80 & 1 \\
3 & Q0933+733 & 2.528  & Quiet  & $-28.0$  & $<$ 0.21 & 1.78 & 1 \\
4 & Q1413+117 & 2.551  & Quiet (BAL) & $-28.3$  & 0.42 & 1.54 & 1 \\
5 & Q0636+680 & 3.195  & Loud (Flat) & $-29.8$  & $<0.1^f$ & 2.3$^f$ & 2 \\
6 & Q0014+813 & 3.398  & Loud (Flat) & $-30.2$  & $<0.6^f$ & 2.3$^f$  & 2 \\
7 & B1422+231 & 3.620  & Loud (Flat) & $-29.2^g$  &  0.26  & 0.16  & 3 \\
8 & PKS1937-101 & 3.787  & Loud (Flat) & $-28.8$  & 0.36  & $<0.49^h$ & 4 \\
\hline \hline
\end{tabular}
\end{center}


\tablenotetext{a}{
Loud = Radio loud, Flat = flat-spectrum source,  Quiet =
radio quiet, BAL = broad-absorption-line feature.}
\tablenotetext{b}{
The absolute $V$ magnitude. A Hubble constant, $H_0 = 100 $ km s$^{-1}$ Mpc$^{-1}$, and
 a deceleration parameter,
$q_0 =0$, are assumed.}
\tablenotetext{c}{
The flux ratio. [OIII] = [OIII]$\lambda$4959 + [OIII]$\lambda$5007.}
\tablenotetext{d}{
The flux ratio. FeII = FeII$\lambda$4570 = FeII$\lambda$4434-4684.}
\tablenotetext{e}{
References: 1. Hill et al. (1993), 2. Elston et al. (1994), 3. Kawara 
et al. (1996), 4. This paper.}
\tablenotetext{f}{
Rough estimates by us from the spectra given in Elston et al. (1994).}
\tablenotetext{g}{
Gravitationally amplified. Possible magnification factor is 15 - 30 and thus 
the absolute magnitude
should be reduced by this factor (Hogg \& Blandford 1994; Kormann, Schneider,
\& Bartelmann 1994).}
\tablenotetext{h}{
Upper limit (3$\sigma_{\rm rms}$).}

\tablenum{2}
\caption{A summary of NIR spectroscopy of high-z quasars \label{tbl-2}}

\end{table*}


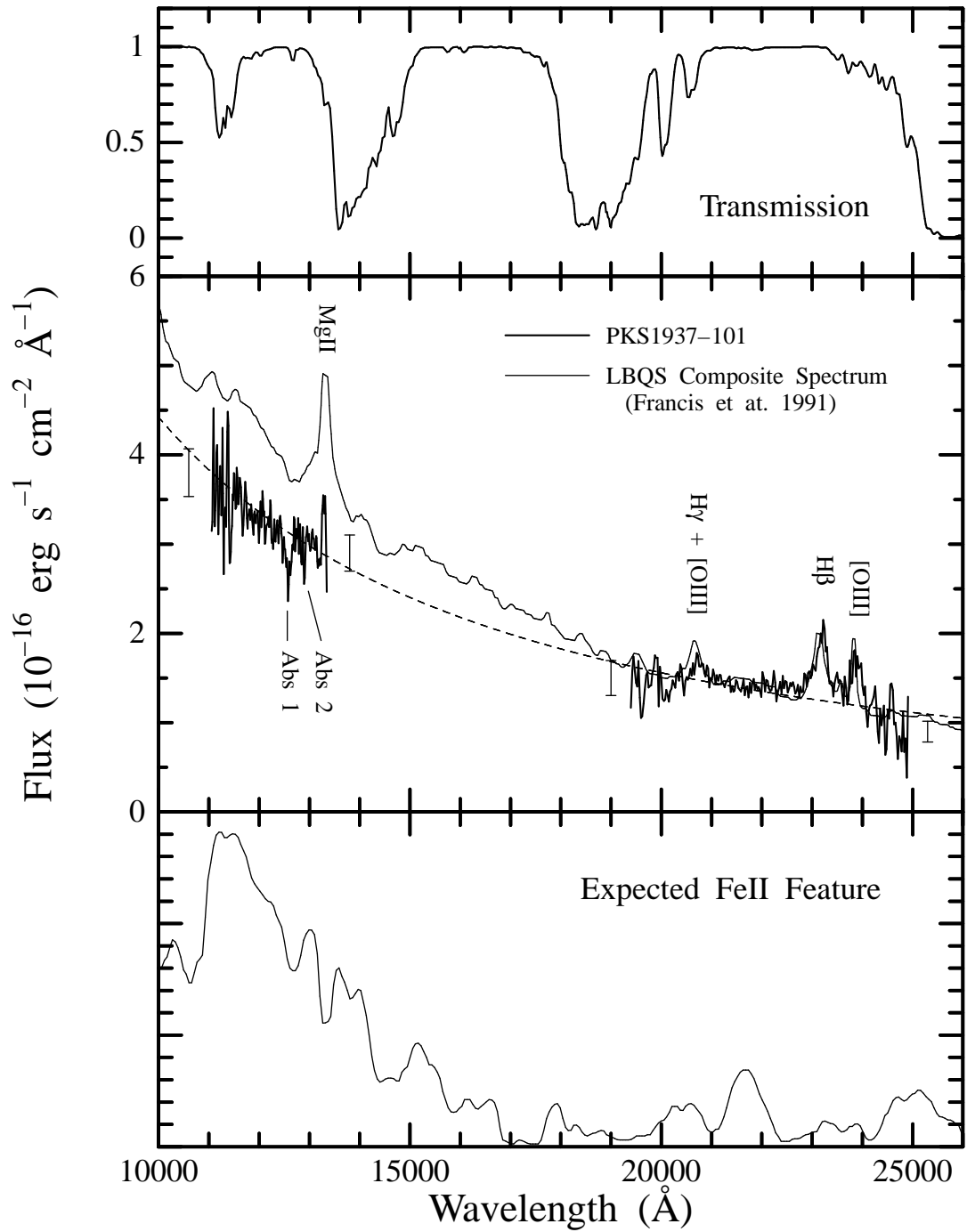
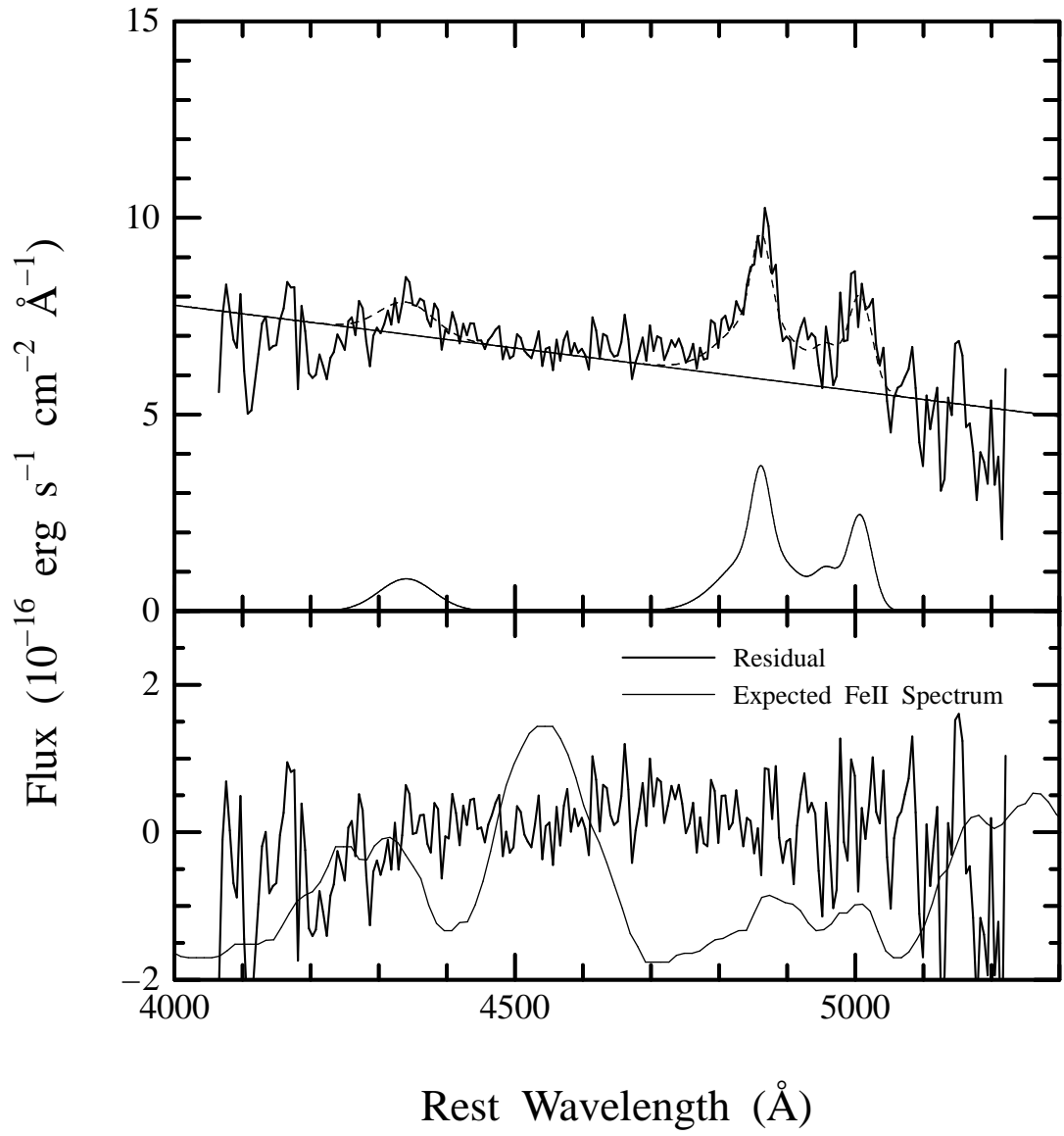
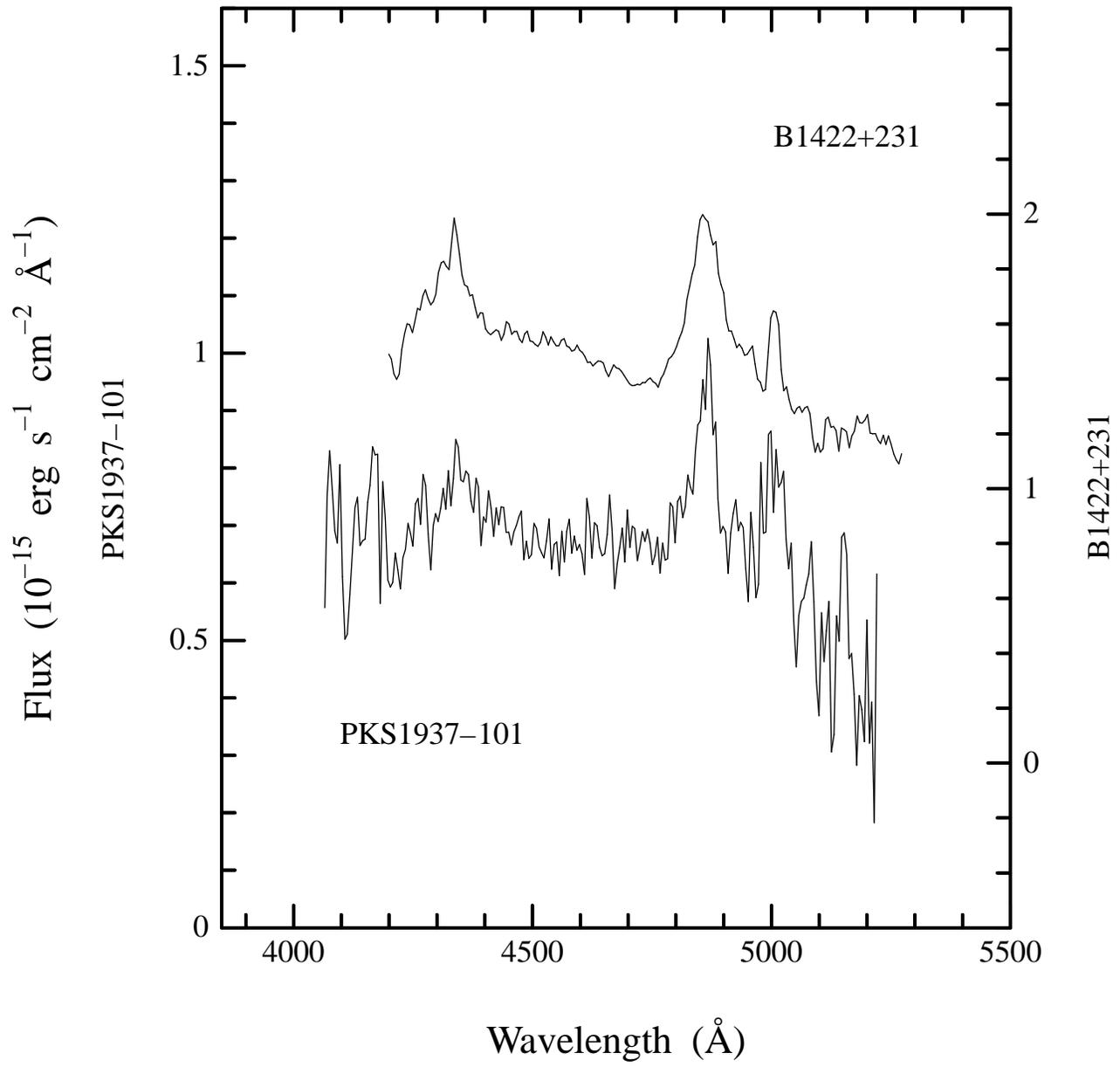
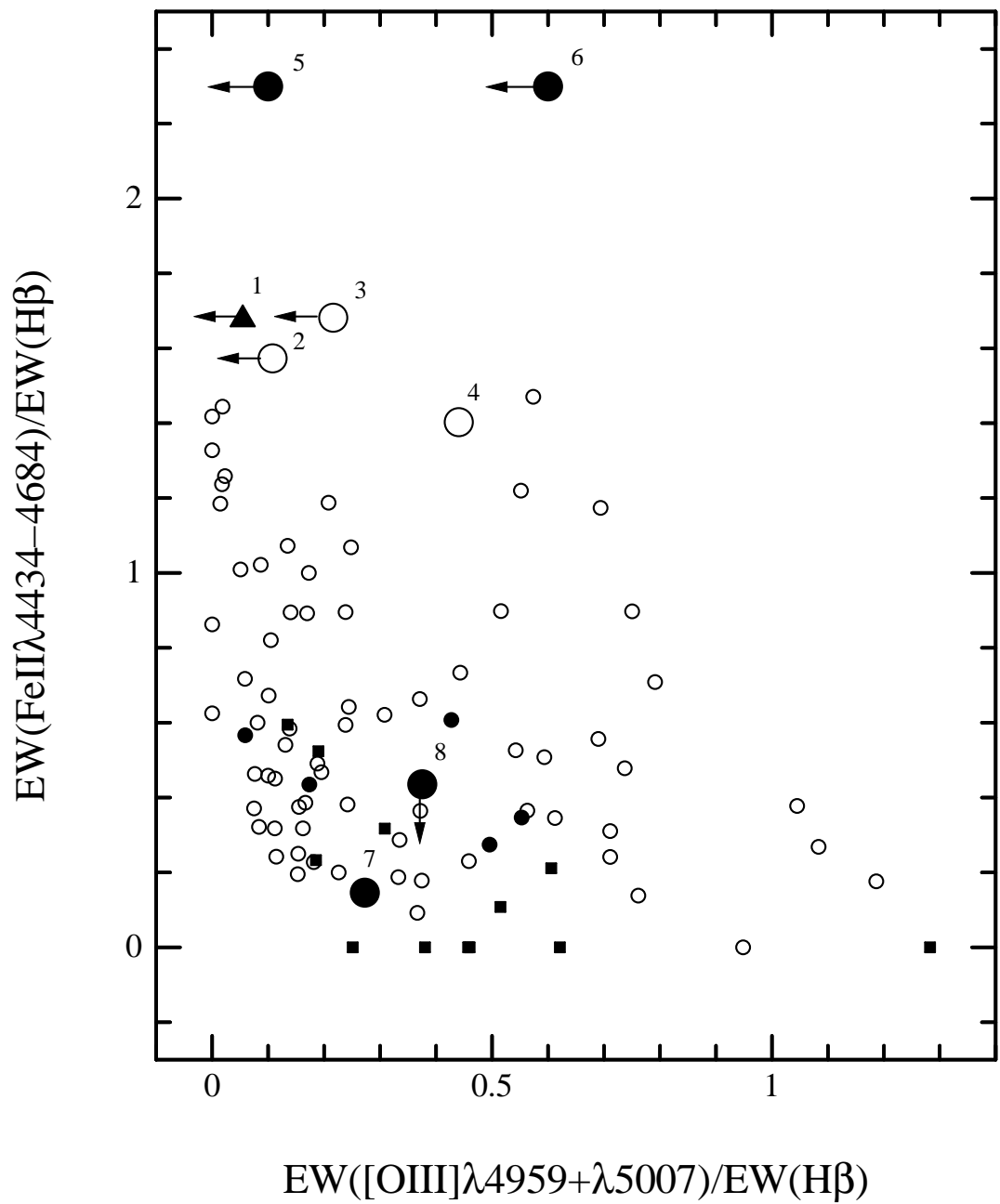
\end{document}